\documentclass[english,prl,groupedaddress,reprint,longbibliography]{revtex4-1}
\usepackage[utf8]{inputenc}
\usepackage{amsmath}
\usepackage{amsfonts}

\usepackage[normalem]{ulem}
\usepackage{amssymb}
\usepackage{braket}
\usepackage{graphicx}
\usepackage{float}
\usepackage{comment}
\usepackage{xcolor}
\usepackage[colorlinks=true,
   linkcolor=blue,
   citecolor=blue,
   urlcolor =blue]{hyperref}

\definecolor{darkpastelgreen}{rgb}{0.01, 0.75, 0.24}
 
\usepackage[normalem]{ulem}
\usepackage{bbm}

\begin{document}
\title{Quantum information scrambling in a trapped-ion quantum simulator with tunable range interactions}
	\author{
	Manoj K. Joshi\textsuperscript{1,2},
	Andreas Elben\textsuperscript{1,2},
	Beno\^it Vermersch\textsuperscript{1,2,3},
	Tiff Brydges\textsuperscript{1,2}, Christine Maier\textsuperscript{1,2}, Peter Zoller\textsuperscript{1,2}, Rainer Blatt\textsuperscript{1,2}, Christian F. Roos\textsuperscript{1,2}}
		\affiliation{\textsuperscript{1}{Center for Quantum Physics, University of Innsbruck, Innsbruck A-6020, Austria,}}
	\affiliation{\textsuperscript{2}{Institute for Quantum Optics and Quantum Information of the Austrian Academy of Sciences,  Innsbruck A-6020, Austria.}}
\affiliation{\textsuperscript{3}{Univ.   Grenoble  Alpes,  CNRS,  LPMMC,  38000  Grenoble,  France}}
\date{\today}
\begin{abstract}
In ergodic many-body quantum systems, locally encoded quantum information becomes, in the course of time evolution, inaccessible to local measurements. This concept of ``scrambling'' is currently of intense research interest, entailing a deep understanding of many-body dynamics such as the processes of chaos and thermalization. Here, we present first experimental demonstrations of quantum information scrambling on a $10$-qubit trapped-ion quantum simulator representing a tunable long-range interacting spin system, by estimating out-of-time ordered correlators (OTOCs) through randomized measurements. We also analyze the role of decoherence in our system by comparing our measurements to numerical simulations and by measuring  R\'enyi entanglement entropies.
\end{abstract}
\maketitle

Synthetic quantum systems of atoms, ions and superconducting qubits provide us with excellent platforms for studying fundamental aspects of the quantum information dynamics~\cite{Lewis-Swan2019}. Existing tabletop experiments with high fidelity quantum control have great prospects in exploiting features of quantum dynamics related to black holes and gravity models, high energy physics models and condensed matter systems \cite{Landsman2018,Muschik2017, kokail2019}. These platforms can probe essential out-of-equilibrium phenomena of interacting many-body systems, such as quantum chaos, thermalization and many-body localization~\cite{DAlessio2016, Eisert2015}. In particular, the systems with single-site control have demonstrated the spreading of time-ordered correlations~\cite{Cheneau2012,Jurcevic2014,Richerme2014} and they have confirmed the existence of Lieb-Robinson bounds~\cite{Lieb1972} in non-relativistic locally interacting systems~\cite{Hauke2013, Schachenmayer2013}.

Recently, a novel concept has been shown to have the ability to identify a fundamental feature of many-body quantum dynamics: quantum information scrambling.
Here, ``scrambling'' describes how quantum information, initially encoded in terms of local operators, becomes after time-evolution increasingly non-local and complex \cite{Shenker2014}.
As a quantum version of the butterfly effect \cite{Lorenz1972},  scrambling can be identified with the decay of a new type of many-point correlation functions, namely, \emph{out-of-time ordered correlations} (OTOCs) \cite{Sachdev1993,Kitaev}. OTOCs have been first used to quantify  the  properties of ``fast scrambling'' governed by universal Lyapunov exponents in the dynamics of black  holes~\cite{Sachdev1993,Kitaev,Hayden2007,Shenker2014,Maldacena2016,Hosur2016,Roberts2017,Banerjee2017}, and to describe chaotic systems with holographic duals~\cite{Roberts2016,Blake2016}.
For quantum lattice models, which are more relevant to experiments with current quantum simulators, OTOCs have the form
\begin{align}
O(t)= D^{-1}\mathrm{Tr}\left[ W(t)V W(t)V \right],
\label{eq:OTOC}
\end{align}
with  $W(t) = e^{iHt} W e^{-iHt}$ a time-evolved Heisenberg operator,  $W$ and $V$  local, hermitian operators, $H$ a many-body Hamiltonian and $D$ the Hilbert space dimension~\footnote{Note, that in general $O(t)= \mathrm{Tr}\left[\rho W(t)V W(t)V \right]$ for some state $\rho$. Here, we consider the infinite temperature OTOC $\rho=\mathbbm{1}/D$.}. OTOCs have been shown to detect universal signatures of many-body quantum chaos~\cite{Bohrdt2017,Nahum2018,VonKeyserlingk2018}, many-body localization~\cite{Fan2016,Huang2017,Chen2017}, and dynamical quantum phase transitions~\cite{Heyl2018}.
In particular, in ergodic systems with local interactions, the decay of the OTOCs of local operators $W$ and $V$ initially separated by a distance $d$ occurs at a characteristic time $t_c\sim d/v_B$, where $v_B$ is the butterfly velocity \cite{Hosur2016}. This result can be understood from a hydrodynamical description associated with a {\it ballistic} spatial spreading of the operator $W(t)$, whose ``wavefront'' travels with velocity $v_B$ and also broadens diffusively in time~\cite{Bohrdt2017,Nahum2018,VonKeyserlingk2018,Xu2019,Couch2019}.
While in seminal experimental work, OTOCs have been measured  either for {\em collective}  (non-local) operators ~\cite{Garttner2017, Wei2018}  or in small-scale $3$-$4$ qubit  systems~\cite{Li2017,Landsman2018}, scrambling  quantified by the OTOCs of local operators (encoding local  quantum information) has so far eluded observation in a many-body system.

In this letter, we present first measurements of OTOCs in a spin-model consisting of $N=10$ qubits {\em  with local interactions of tunable range}, realized in a trapped-ion quantum simulator. To this end, {we} implement a recently proposed protocol, based on measuring statistical correlations of randomized measurements~\cite{Vermersch2019}, to access OTOCs as defined in Eq.~(\ref{eq:OTOC}). This allows us in particular to monitor the emergence of a traveling operator wavefront, and observe the crucial role played by the interaction range
~\cite{Foss-Feig2015,Matsuta2017,Tran2019,Else2018,Luitz2019,Zhou2019}.
Furthermore, we demonstrate the robustness of the implemented protocol against certain types of decoherence mechanisms. Most importantly, the implemented protocol neither relies on time-reversed evolution nor auxiliary degrees of freedom required in previous proposals~\cite{Swingle2016,Yao2016,Zhu2016,Swingle2018,Yoshida2019,GonzalezAlonso2019}. To verify its robustness, we compare our measurement results with numerical simulations, and  perform additional randomized measurements of operator spreading~\cite{Qi2019}, and entanglement R\'enyi entropies~\cite{Elben2018,Brydges2019}. 

\begin{figure}[hbt!]
    \centering
    \includegraphics[scale=0.42]{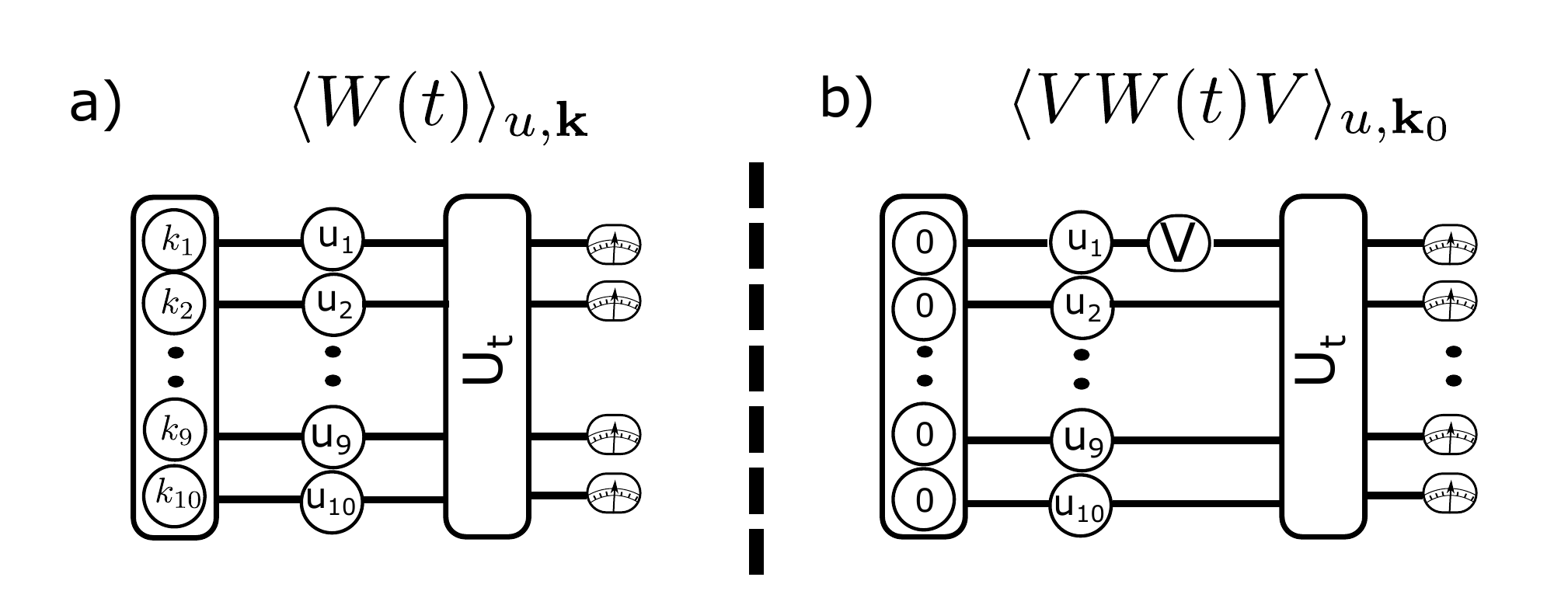}\\
     \includegraphics[scale=1.08]{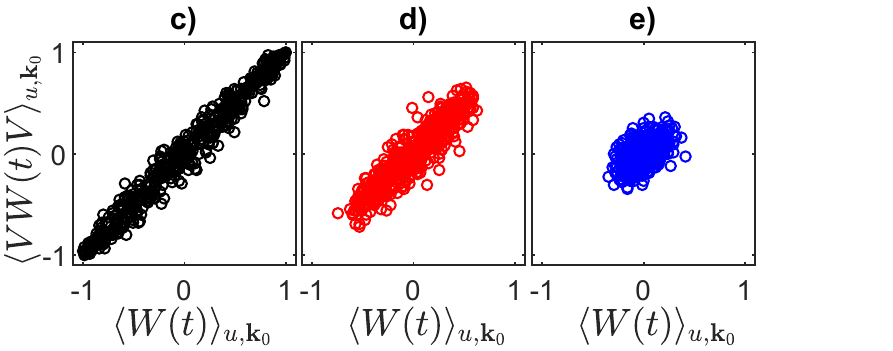}
    \caption{a-b) Experimental procedure to measure two parts of the OTOCs $\braket{W(t)}_{u,\mathbf{k}}$ and $\braket{VW(t)V}_{u,\mathbf{k}_0}$. Here, $\mathbf{k}$ ($\mathbf{k}_0$)  refers to the initial product state $\ket{\mathbf{k}}=\ket{k_1,k_2...k_n}$ ($\ket{\mathbf{k}_0}=\ket{0,\dots,0}$) with $k_i\in \{0, 1\}$. c-e) Spread of operator $W(t)=\sigma_5^z(t)$ is observed in terms of loss of correlations between two measured quantities ($\braket{W(t)}_{u,\mathbf{k}_0}$ and $\braket{VW(t)V}_{u,\mathbf{k_0}}$) for $t=0,2$ and $5$ ms, where $V=\sigma_1^z$. The experimental study is carried out for $\alpha = 1.21$, $J_{0} = 2\pi \times 30.13$ Hz and $B =2\pi\times1.5$ kHz (see SM for the experimental details \cite{SM}). }
    \label{fig:Illustration}
\end{figure}

{\it Experimental setup and protocol}---
The experimental studies are performed on a trapped-ion analog quantum simulator, realized with a linear chain of $N=10$ $^{40}\text{Ca}^+$ ions. Quantum information is encoded in two \mbox{(pseudo-)} spin states $\ket{\text{S}_{1/2},m=+1/2}$ $\equiv \ket{0}$ and $\ket{\text{D}_{5/2},{m=+5/2}}$  $\equiv \ket{1}$, respectively. Entangling operations are performed using a bichromatic laser field, which off-resonantly couples the electronic and vibrational states of the ions \cite{Jurcevic2014}. The resulting interaction Hamiltonian is expressed as
\begin{equation}
    H=\sum_{i \ne j}\frac{J_{0}}{|i-j|^{\alpha}}\sigma_i^x\sigma_j^x + B\sum_i\sigma_i^z,
    \label{eq:Jijinteraction}
\end{equation}
where $i$,$j$ are the indices representing the position of ions in the chain,
and $\sigma^\beta$, $\beta\in\{x,z\}$, denote Pauli spin matrices;
$J_{0}$ and $\alpha$ represent the maximum strength and exponent of the equivalent Ising-type interaction, respectively. Note that the above interaction Hamiltonian leads to a  spin \textit{flip-flop} type interaction when the transverse field $B \gg J_{0}$ (i.e. $H=\sum_{i \ne j}J_{ij}\sigma_i^+\sigma_j^- + B\sum_i\sigma_i^z$), which is routinely used in analog quantum simulators \cite{Jurcevic2014, Richerme2014}. Coherent control of the spin state of an individual qubit is achieved by a tightly focused, steerable laser beam, enabling the preparation of any desired product state $\ket{\Psi_k} =\bigotimes_{i = 1}^{N} \ket{\psi_i}$. More details about the experimental platform can be found in the Supplemental Materials (SM) \cite{SM}\nocite{Blume:2013,Sandia,Blume:2017,Barton2000}. 

The experimental protocol to measure OTOCs consists of two main parts and is illustrated in Fig.~\ref{fig:Illustration}~a) and b). In the first part [Fig.~\ref{fig:Illustration}~a)], we prepare an initial product state $\ket{\mathbf{k}_0}=\ket{0,0,\dots,0}$.  Next, we apply a local random unitary  $u=u_1\otimes \dots \otimes u_N$ where each $u_i$, implementing a random single spin rotation,  is sampled independently from the circular unitary ensemble (CUE) \cite{Mezzadri2006}. As investigated in detail in Refs.~\cite{Brydges2019, Elben2020}, these unitaries are generated with high fidelity in our apparatus (see SM for single-qubit gate fidelity \cite{SM}). Subsequently, the system is evolved in the presence of the Ising Hamiltonian $H$, for time $t$ and  the operator $W=\sigma_j^x$, for $j\in \{1,\dots,N\}$, is measured. After  $N_M=150$ ($N_M=300$ for $t=4,5$ ms) repetitions, one obtains an estimation of $\langle W(t) \rangle_{u,\mathbf{k}_0}=\bra{\mathbf{k}_0} u^\dag W(t) u\ket{\mathbf{k}_0}$. In the second part of the experiment, we prepare the  initial product  state $\ket{\mathbf{k}_0}$ and repeat the  experimental procedure {with the same random unitary $u$}. In this part, a unitary $V=\sigma_1^z$, is applied, in addition, before the time evolution [see Fig.\ref{fig:Illustration} b)]. By repetition ($N_M=150$ or $N_M=300$), we obtain an estimation of $\langle V W(t) V \rangle_{u,\mathbf{k}_0}$. The steps are illustrated in Fig.~\ref{fig:Illustration}~b). Both parts are finally repeated for $N_U=500$ sets of unitaries $u$ to build statistical correlations.  Note that the choice of $N_M$ and $N_U$ determines the expected statistical error which is investigated in detail in Ref.~\ \cite{Vermersch2019}.

The basic intuition to understand how statistical correlations between two randomized measurements $\langle W(t) \rangle_{u,\mathbf{k}_0}$ and $\langle V W(t)V \rangle_{u,\mathbf{k}_0}$ can be used as probes for operator spreading, and how they are related to OTOCs is provided in Fig.~\ref{fig:Illustration}~c-e). The figure shows experimentally measured expectation values $\langle W(t) \rangle_{u,\mathbf{k}_0}$ and $\langle V W(t)V \rangle_{u,\mathbf{k}_0}$ for $N_U=500$ unitaries $u$, $V=\sigma_1^z$ and $W=\sigma_5^x$. At initial time $t=0$~ms [panel c)], the system has not evolved under $H$, and so the measurement of $W$ at $j=5$ is not affected by whether $V$ has been applied at $j=1$ or not. Thus, we observe (up to projection noise) near perfect correlations between $\langle W(t) \rangle_{u,\mathbf{k}_0}$, and $\langle V W(t)V \rangle_{u,\mathbf{k}_0}$.  At later times $t=2$~ms and $t=5$~ms, as shown in Fig.\ \ref{fig:Illustration}, panels d) and e), the  information that $V$ had been applied at $j=1$ has spread over the system, and hence the measurement of $W$ at $j=5$ is affected. As an effect, the correlations between $\langle W(t) \rangle_{u,\mathbf{k}_0}$, and $\langle V W(t)V \rangle_{u,\mathbf{k}_0}$ decrease with time, as the OTOCs in ergodic systems.

The formal mapping of  correlations between expectation values  obtained via  forward time-evolution from randomized initial states  and out-of-time-ordered correlation functions has been derived in Ref.\ \cite{Vermersch2019}. For the local random unitaries employed here, the first part  of the protocol  [Fig.~\ref{fig:Illustration}~a)] is to this end  repeated (with the same unitaries $u$) for a set $E_n=\{\mathbf{k}_0,\dots,\mathbf{k}_{2^n-1}\}$  ($n=2$ in the context of this work, see below) of initial product states  $\mathbf{k} \in E_n$ to obtain estimations of $\langle W(t) \rangle_{u,\mathbf{k}}=\bra{\mathbf{k}} u^\dag W(t) u\ket{\mathbf{k}}$ for all $\mathbf{k} \in E_n$.   Here,  $\mathbf{k}_s=(k_{s,1},\dots,k_{s,N}) $ with $k_{s,i} \in \{0,1\}$ is given by  the reverse $N$-bit binary representation of $s$, e.g.\ $\mathbf{k}_0=(0,0,\dots,0)\,,\, \mathbf{k}_1=(1,0,\dots,0)\, ,\, \mathbf{k}_2=(0,1,0,\dots,0)\,,\, \mathbf{k}_3=(1,1,0,\dots,0)$ \cite{Vermersch2019}. The summed correlations $ \overline{\langle W(t) \rangle_{u,\mathbf{k}} \langle VW(t)V \rangle_{u,\mathbf{k}_0}} $ for all $\mathbf{k} \in E_n$ map then to  `modified' OTOCs \cite{Vermersch2019}
 \begin{align}
\label{eq:defOLn}
O_n(t) 
&=   \frac{\sum_{\mathbf{k}\in E_n}  (-2)^{-D[\mathbf{k}_0,\mathbf{k}]} \overline{\langle W(t) \rangle_{u,\mathbf{k}} \langle  V W(t) V  \rangle_{u,\mathbf{k}_0}}}{\sum_{\mathbf{k}\in E_n} (-2)^{-D[\mathbf{k}_0,\mathbf{k}]} \overline{\langle W(t) \rangle_{u,\mathbf{k}} \langle W(t)   \rangle_{u,\mathbf{k}_0}}}, 
\end{align}
for $n\in \{0,\dots, N\}$ and $\overline{\vphantom{H}\dots}$ the ensemble average over the random unitaries $u$. Here, $D[\mathbf{k}_0,\mathbf{k}]$ is the Hamming distance between $\mathbf{k}_0$ and $\mathbf{k}$, i.e. the number of spin flips to transfer $\ket{\mathbf{k}_0}$ into $\ket{\mathbf{k}}$. As shown in Ref.\cite{Vermersch2019} and SM \cite{SM}, the modified OTOCs $O_n(t)$, with $n=0,1 \dots N$, represent a series with fast convergence to $O(t)$ with increasing $n$, in particular $O_N(t)=O(t)$. 
This means that, to obtain a quantitative measure of the OTOC $O(t)$, the modified OTOCs $O_0(t)$, $O_1(t)$,\dots should be measured (with increasing experimental effort) until convergence to $O(t)$. These convergence aspects are discussed in details in Ref.\cite{Vermersch2019} in the context of several physical examples.
In our case, we obtain approximate convergence to $O(t)$ at $n=2$ (see below and SM). Note that the lowest order modified OTOC, $O_0(t)$, has been measured using this method in a four qubit NMR system~\cite{Nie2019}. 

\begin{figure}[t]
    \centering
    \includegraphics[scale=0.9]{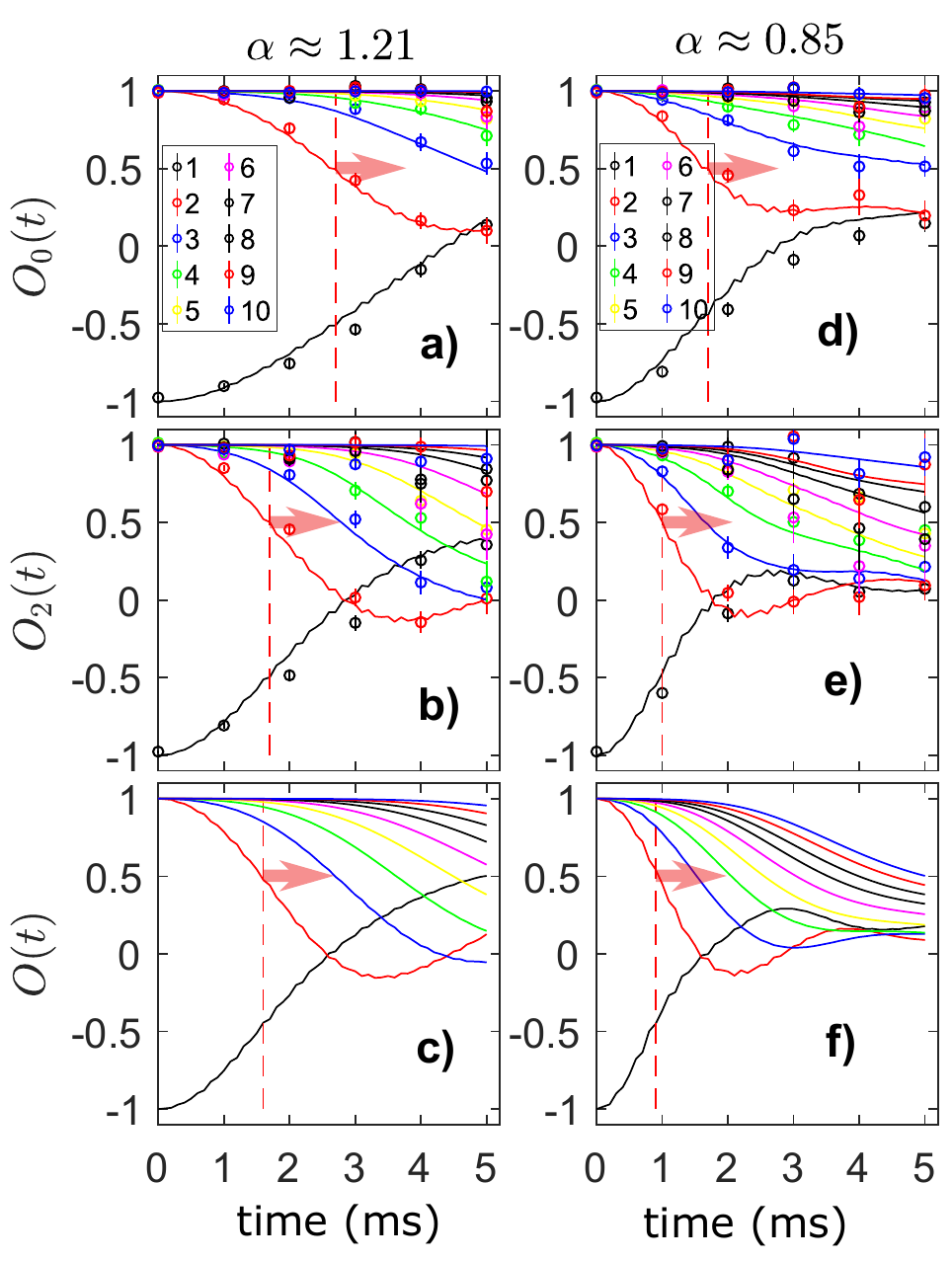}
     \caption{ Out-of-time ordered correlators in a 10-qubit quantum simulator for local operators $V=\sigma_1^z$ and $W(t)=\sigma_j^x(t)$ for $j=1,2,\dots,10$ (color coded in the figure). On the left panel, estimated OTOCs namely $O_0(t)$ in a), and $O_2(t)$ in b) are shown for various interaction times and qubits at interaction exponent $\alpha = 1.21$ and  $J_0 = 2\pi \times 30.13$ Hz and $B =2\pi\times1.5$ kHz. On the right panel, measurement of OTOCs  at $\alpha = 0.85$, $J_{0} = 2\pi \times 40.78$ Hz and $B =2\pi\times1.5$ kHz. Here circles are experimental points and lines are numerical results for the experimental parameters. c) and f) are the exact OTOCs $O(t)=O_{10}(t)$ simulated for the aforementioned parameters. Error bars associated with the experimental measurements are of the size of the symbols and they are deduced by the Jackknife sampling method. Here, red arrows indicate direction of propagation of the \emph{operator wavefront} and corresponding vertical dashed lines indicate the times when OTOCs decay to 0.5 (see main text for a detailed discussion). } 
    \label{fig:OTOCbothalpha}
\end{figure}

{\it Measurement of OTOCs ---}
We now present measurements of the modified OTOCs $O_n(t)$ for $n=0,2$ for two values of the power law exponent $\alpha=1.21$ (long-range interaction), and $0.85$ (corresponding to a very long-range interaction) and demonstrate the fast convergence to $O(t)$ by a comparison to numerical simulations. For our Hamiltonian evolution with long-range interactions, the operator wavefront is not expected to spread in a purely ballistic manner and the shape of the spatial-temporal profile of time ordered~\cite{Hauke2013,Schachenmayer2013} and out-of-time ordered correlations~\cite{Foss-Feig2015,Matsuta2017,Tran2019,Else2018,Luitz2019,Zhou2019} is the matter of current theoretical investigations. In Fig.~\ref{fig:OTOCbothalpha} a)-b), the measured OTOCs $O_0(t)$, $O_2(t)$ (circles) are plotted as a function of time $t$ after the quantum quench at $\alpha = 1.21$, which we compare with numerical simulations (solid lines) assuming unitary time evolution. The error bars are obtained via the Jackknife method \cite{Shao2012}. The exact OTOC $O(t)$ calculated from Eq.~\eqref{eq:defOLn} is shown in panel c).
All two measured OTOCs display the same qualitative behavior; initially near-perfect (anti-) correlations exist for measurements of $W=\sigma_j^x$ performed at ion $j>1$ ($j=1$, respectively) and hence reveal spatio-temporal profiles of the OTOCs in the long-range interacting system. This is described as wavefront propagation of local perturbation from the causal site, i.e. the site at which $V(0)$ operator is encoded, to the effect site where the operator $W(t)$ is measured. For the current studies, the propagation of the wavefront is indicated with an arrow in Fig. \ref{fig:OTOCbothalpha}. For $\alpha= 0.85$, Fig.~\ref{fig:OTOCbothalpha} right panel, corresponding to even longer range interactions than the aforementioned case,  the dynamics of OTOCs look qualitatively different compared to $\alpha = 1.21$. Particularly, here, the dynamics are faster than in the former case. A detailed discussion of the quantitative differences is given below.  

At the quantitative level, two observations are apparent: (i) experimentally measured and theoretically simulated OTOCs are, within error limits of the experiment (see SM \cite{SM}), in good agreement, implying consistency of the protocol while measuring OTOCs in our system. Furthermore, since the theoretical curves are obtained by simulating unitary dynamics, this demonstrates that the measurement protocol is not affected by decoherence, which appears due to global dephasing in the experiment (see below and SM for further discussion \cite{SM}). (ii) $O_0(t)$ describes the same qualitative behavior as $O_2(t)$ but quantitatively differs from the actual OTOC $O(t)$, thus corroborating poor approximations of the OTOCs as predicted by the theory \cite{Vermersch2019}. On the contrary, the OTOC $O_2(t)$ provides a good approximation to $O(t)$, and captures in particular the features of the operator spreading [compare panel b), and c), and panel e) and f)]. This means that the sampling procedure described above to access the converging series $O_n(t)$ is adapted to our experimental setup \cite{SM}. Furthermore, the OTOC measurements slightly deviate at the later times of $t=4,5$~ms. This deviation might be due to uncertainties in the determination of the Hamiltonian parameters, which are estimated through measurements of local excitation spread in the ion chain~\cite{Jurcevic2014}. For further details, see the SM \cite{SM}.

\begin{figure}[t]
	\centering
	\includegraphics[scale=0.75]{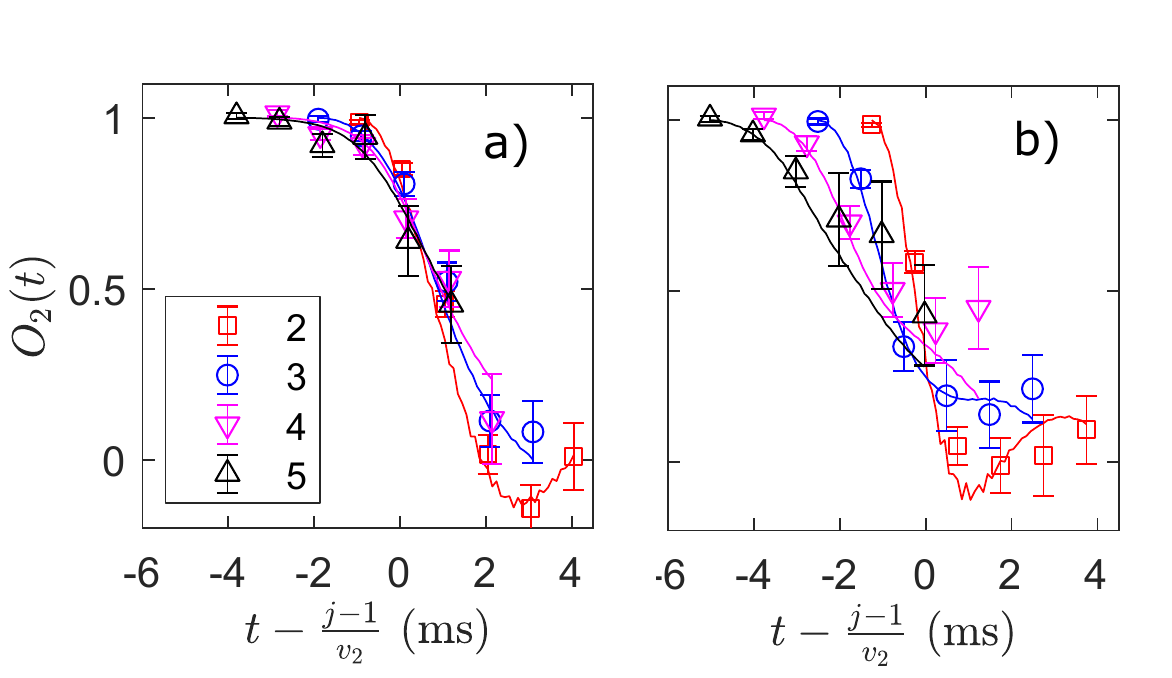}
 \caption{Out-of-time-ordered correlators versus rescaled time for interactions with power-law exponents a) $\alpha = 1.21$ and b) $\alpha = 0.85$ and for $W(t)=\sigma_j^x(t)$ located at spins $j=2,3,4,5$ (red, blue, pink, black). For a)  $v_2= (1.0 \pm 0.2) \!\cdot \! 10^3 \,\textrm{s}^{-1}$  and b) $v_2= (0.8 \pm 0.2) \! \cdot \!10^3 \,\textrm{s}^{-1}$ are fitted such that we observe the best possible collapse of the data at a threshold value $0.5$ (see SM for details \cite{SM}).
 }
	\label{fig:collapseplot}
\end{figure}

In Fig.~\ref{fig:collapseplot} a) and b), we study the shape of the spatio-temporal profile of the OTOCs. To this end, we rescale the time axis $t-(j-1)/v_2$, with $v_2$ chosen such that we observe the best possible collapse of the measured data for various locations $j$ of $W$ at a threshold value of $O_2=0.5$ (see SM for details of the fitting procedure \cite{SM}). In Fig.~3a), for a power-law exponent $\alpha = 1.21$, we find that the measured data indeed collapses.  Our early time data is thus consistent with a ballistic expansion of the operator wavefront, with  velocity $v_2=(1.0 \pm 0.2) \!\cdot \! 10^3 \,\textrm{s}^{-1}$. 
We note that, due to finite time and size effects,  a slow emergence of super-ballistic behavior,  predicted for large systems and  $\alpha = 1.2$~\cite{Zhou2019}, cannot be unambiguously distinguished with our experimental data. In Fig.~3b), for a power-law exponent $\alpha=0.85$, we do not observe a collapse of the measured data to a single curve, meaning that the shape of the operator wavefront is not conserved over time and space, and therefore that the dynamics is not ballistic. A broadening of the decay of $O_2$ with time and distance is instead clearly visible. While we emphasize that the spatio-temporal profiles still show some differences when comparing $O_2(t)$ and the exact OTOC $O(t)$, c.f Fig.~\ref{fig:OTOCbothalpha}, this strong broadening is consistent with the theoretical prediction for $\alpha<1$ \cite{Zhou2019}.

{\it  Other probes of the scrambling of quantum information---} 
In generic quantum systems,  the scrambling of quantum information does not only manifest itself through the decay of the OTOCs but also through a decrease of statistical moments of the type $\overline{\langle W(t) \rangle^2 _{u,\mathbf{k_0}}}$~\cite{Qi2019}, and an increase of entanglement entropies~\cite{Lewis-Swan2019}. As we show now,  the measurement of these two quantities, which are both accessible via randomized measurements,  provides us both with evidence of operator spreading that are complementary to OTOCs, and allows us to identify and assess the role of decoherence in our experiment.

\begin{figure}[t]
    \centering
  \includegraphics[scale=1.5]{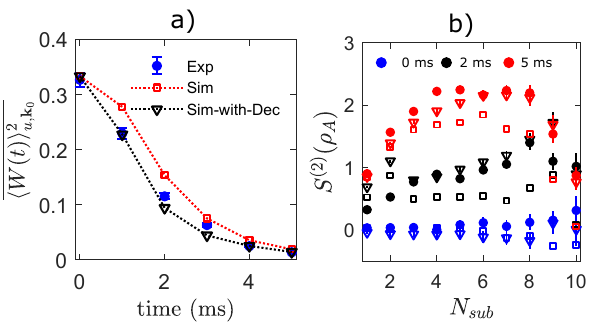}
    \caption{a) Time evolution of $\overline{  \braket{W(t)}_{u,\mathbf{k}_0}^2}=\overline{  \braket{\sigma_j^x(t)}_{u,\mathbf{k}_0}^2}$,  averaged over $j$ (all qubits 1 to 10). Squares (red) and triangles (black) are numerical simulations of unitary dynamics and including decoherence effects, respectively. 
    Dashed lines are guides to the eye.
    b) Additionally, R\'enyi entropy measurements (in circles) are carried out at $t=0$ (blue), 2 ms (black) and 5 ms (red) for partitions of the form $A=\{1,\dots,N_\mathrm{sub}\}$. Squares and triangles are theoretical simulations without and with decoherence, respectively. The experimental and simulation parameters for studies presented in (a) and (b) are $\alpha$ = 1.21, $J_0 = 2 \pi \times 30.13 $ Hz and $B = 2 \pi \times 1.5$ kHz.  }
    \label{fig:WsquareandPurities}
\end{figure}

The second moment of the expectation value $\overline{\langle W(t) \rangle^2 _{u,\mathbf{k_0}}}$ is  accessed from statistical \emph{auto}-correlations of randomized measurements performed on a single system. Its measurement is enabled through the first part of the OTOC protocol [Fig.~\ref{fig:Illustration} a)]. Note that $\overline{\langle W(t) \rangle^2 _{u,\mathbf{k_0}}}$  appears also as normalization in the denominator of Eq.~\eqref{eq:defOLn}. As shown in Fig.~\ref{fig:WsquareandPurities}a), $\overline{\langle W(t) \rangle^2 _{u,\mathbf{k_0}}}$  decreases with time, providing, for unitary dynamics, a direct signature of operator spreading and scrambling \cite{Qi2019}. We emphasize that decoherence has a small  decreasing effect on the measurement results. This is due to the fact that decoherence drives the system towards a steady state with reduced magnetization. Hence, both, decoherence and scrambling lead to a decay of  $\overline{\langle W(t) \rangle^2 _{u,\mathbf{k_0}}}$ with time. In contrast, the OTOC measurement is  not affected by decoherence, because our estimation from Eq.~\eqref{eq:defOLn} is normalized (see Figs.~\ref{fig:OTOCbothalpha}-\ref{fig:collapseplot}, for the comparison to unitary theory, and SM for simulations with decoherence \cite{SM}).
 
R\'enyi entanglement entropies quantifying bipartite entanglement are directly related to universal properties of operator spreading~\cite{Lewis-Swan2019,VonKeyserlingk2018,Nahum2018}, and allow us to observe  direct effects of decoherence. The growth of R\'enyi entanglement entropy was previously measured in Ref.~\cite{Brydges2019} where the effects of decoherence were suppressed by starting the quantum quench from an initial N\'eel state in a decoherence-free subspace. In contrast, the effects of decoherence can be made visible for an initial condition as in the OTOC measurements, by choosing a random initial state $\bigotimes_{j=1}^N u_i\ket{\mathbf{k}_0}$ (with fixed local random unitaries $u_i$). This state is not protected against decoherence in time evolution. We evolve this state under $H$, and we measure the second R\'enyi entropy of the final state, and of reduced density matrices of arbitrary partitions $A$ following Ref.~\cite{Brydges2019,Elben2019}. Fig.~\ref{fig:WsquareandPurities}~b) shows an increase of the entropy of the total system to around $S^{(2)}(\rho_A)=-\log_2 \textrm{Tr}[\rho_A^2]= 0.8$ at $t=2$~ms and $t=5$~ms, signaling the presence of decoherence, and in quantitative agreement with our numerical simulations \cite{SM}. However, the entropy of the subsystems acquires even higher values, which demonstrates the presence of bipartite entanglement~\cite{Brydges2019} associated with operator spreading.

 {\it Conclusion and outlook---}
We have presented  measurements of out-of-time ordered correlators in a system of trapped ions with power-law interactions of tunable range. We have demonstrated how the ``wavefront'' of a local operator propagates in such systems, leading to spatial delocalization of quantum information, and  scrambling. The key ingredients of the utilized measurement protocol are randomized measurements which can be implemented with current state-of-the-art technology in various synthetic quantum systems. Their usability is not only feasible with trapped ions but also with Rydberg atoms, optical cavity systems, and superconducting qubits, hence advocating for a powerful and generic method to probe quantum dynamics. The ability to access (modified-)OTOCs in various setups, and their convergence to the exact ones, motivates new approaches to engineer various types of quantum dynamics, in particular in the situation of ``fast scrambling'' relevant to  quantum gravity~\cite{Danshita2017,Marino2019,Bentsen2019} and ``out-of-equilibrium" dynamics in lattice systems~\cite{Eisert2015}.

\begin{acknowledgments}
We thank  J.~Bollinger, P.~Hrmo, L.~K.~Joshi, C.~Monroe and his group, F.~Pollmann, A.~Rey, and J.~Ye for discussions, and L.~Sieberer and  N.~Yao for their contributions to the theory proposal. The project has received funding from the European Research Council (ERC) under the European Union’s Horizon 2020 research and innovation programme (Grant Agreement No. 741541), and from the European Union’s Horizon 2020 research and innovation programme under Grant Agreement No. 817482 (Pasquans) and No. 731473 (QuantERA via QTFLAG). Furthermore, this work was supported by the Simons Collaboration on Ultra-Quantum Matter, which is a grant from the Simons Foundation (651440, P.Z.) and LASCEM by AFOSR No.\ 64896-PH-QC. We acknowledge support by the ERC Synergy Grant UQUAM and by the Austrian Science Fund through the SFB BeyondC (F71). Numerical simulations were realized with QuTiP~\cite{johansson2012qutip}.
\end{acknowledgments}
\newpage 
\bibliography{OTOC}

\section{Experimental platform and procedures}\label{subs:expsequence}
A chain of 10 $^{40}$Ca$^+$ ions is trapped in a linear Paul trap. The ions are Doppler-cooled for 3~ms, followed by 6.5~ms of sideband cooling. After preparing the ions in the ground state of motion, the ions are optically pumped into $\ket{\text{S}_{1/2},\text{m} =+1/2}$. A desired initial product state $\ket{\Psi_i}$ is then realised with the help of a tightly focused laser beam  in combination with a global beam inducing the same single-qubit operation on all qubits. This results in arbitrary single qubit gates with fidelity $ > 99\%$. Further details regarding the single qubit gate infidelity can be found in the following section. The random unitaries are then drawn from a circular unitary ensemble (CUE) as discussed in \cite{Mezzadri2006, Brydges2019}. A bichromatic laser beam is used to create the entangling operations of variable range.  Finally, quantum state readout is performed by scattering photons on the S$_{1/2} \leftrightarrow$ P$_{1/2}$ transition, and a spatially resolved fluorescence detection is implemented for detecting the state of individual ions in single-shot measurements. In order to avoid effects of systematic errors and slow variation of the experimental parameters while taking the measurement of $\braket{W(t)}_{u,\mathbf{k}_s}$ for $s =$ 0--3 and $\braket{V W(t) V}_{u,\mathbf{k}_0}$, all the measurements are performed sequentially for a given set of random unitaries $u$.

The range of the spin-spin interaction [as expressed in Eq.~(2), main text (MT)]  is varied by changing the detuning of the bichromatic laser beam from the transverse center-of-mass sideband transition and also by changing the axial trap frequency. For achieving an interaction range corresponding to $\alpha = 1.21$, the centre of mass (COM) mode frequencies are chosen to be $\omega_z =2\pi\times217$ kHz, $\omega_x =2\pi\times2.673$ MHz and $\omega_y =2\pi\times2.640$ MHz. Additionally, the bichromatic laser frequency components are detuned from the transition by 40 kHz. A range corresponding to $\alpha = 0.85$ is achieved using an axial frequency of  $\omega_z =2\pi\times304$ kHz, and by setting the bichromatic detuning to be 30 kHz. Both experimental measurements are performed with a transverse field strength of $B=2 \pi\times1.5$ kHz.

\section{Error analysis and noise in the experimental system} \label{subs:errorAnal}

\textit{Statistical errors}
-- For the analysis of the OTOCs, we consider two main sources of statistical errors; the first due to finite numbers of measurements $N_M$ per random unitary (quantum projection noise) and the second due to finite numbers of random unitaries $N_U$ employed to estimate the ensemble average. A detailed description of the statistical error analysis can be found in Ref. \cite{Vermersch2019}. Note that  the actual statistical uncertainty of the experimental results presented in this work is directly estimated from the obtained data using Jacknife resampling across the random unitaries.

\textit{Source of decoherence}-- Beside these statistical errors, the system suffers from dephasing noise caused by fluctuating magnetic fields in our laboratory. The entire trapping assembly is placed inside a $\mu-$metal shield which greatly suppresses magnetic field noise. By Ramsey spectroscopy, We measure the qubit coherence time in the presence of the laser fields that create the entangling interactions. 
These experiments are performed with only a single ion so that the bichromatic laser field off-resonantly couples the qubit to the ion's transverse motion without creating entanglement between multiple qubits. The experimental parameters are the same as for the 10-ion experiments except for the reduce number of vibrational modes. 

\begin{figure}[]
    \centering
  \includegraphics[scale=0.5]{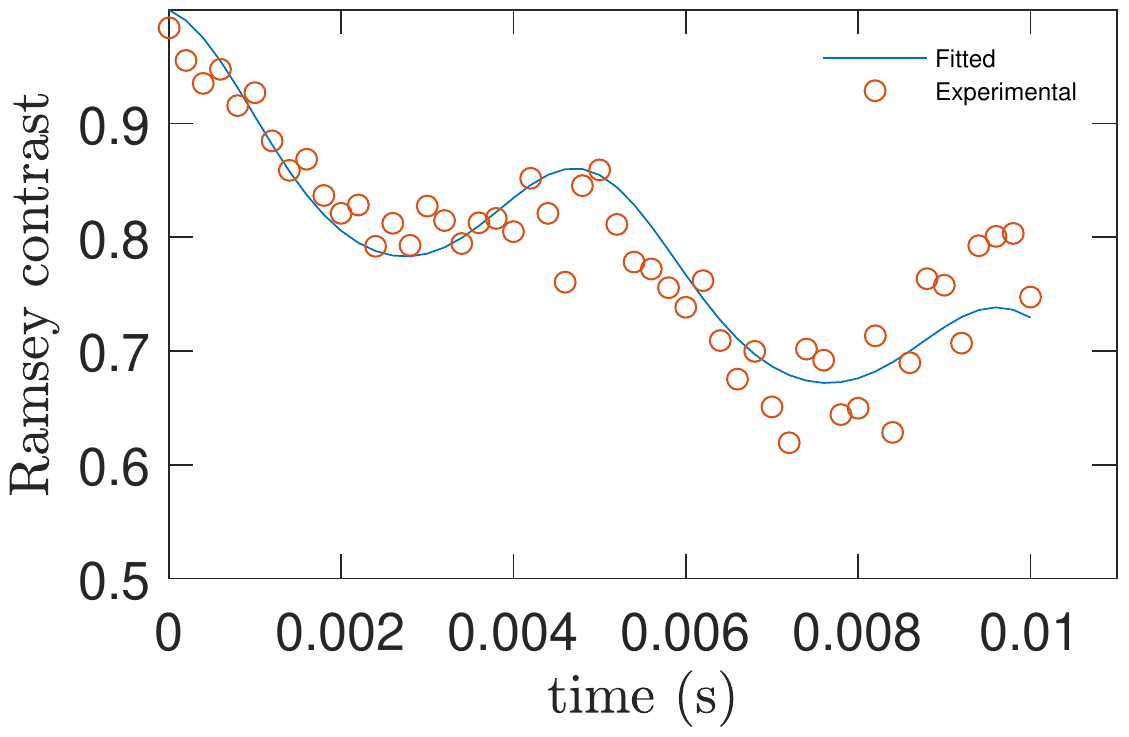}
    \caption{Measurement of coherence time of the qubit via Ramsey spectroscopy. A periodic magnetic field fluctuation is measured at 204 Hz, causing a variation in the transition frequency.}
    \label{fig:decoherence}
\end{figure}

Fig.~\ref{fig:decoherence} shows the experimentally measured Ramsey contrast for variable waiting times between two $\pi/2$ pulses. The bichromatic laser beam is turned on during the Ramsey waiting time such that fluctuating light shifts or incoherent coupling to the ion motion are taken into account. 
Fig.~\ref{fig:decoherence} highlights the loss of Ramsey contrast up to 10~ms. A loss of Ramsey contrast followed by contrast revivals after around 5~ms and 10~ms waiting time 
points to a noise source that is periodically shifting the qubit transition frequency with a frequency of about 200 Hz. It was only after having carried out the OTOC measurements that we identified an unexpected magnetic field noise at 204~Hz, which was subsequently eliminated. For the experiments presented in this paper, we included this noise source in the numerical simulation of OTOC measurements.  After performing least-squares fitting of the Ramsey data in Fig.~\ref{fig:decoherence}, we estimated the amplitude of the periodic noise component to be $2\pi\times90$ Hz. Due to broadband high-frequency magnetic-field and laser phase noise, the contrast decays to $1/e$ after $\tau_{coh}$ = 0.033 s, i.e. a time scale considerably longer than the duration of the probed spin-spin interactions. Nevertheless, the observed non-negligible dephasing was included in the numerical simulations (see the MT and the section below). 

\textit{Errors in single qubit operations}--
Single qubit operations can suffer from several sources of errors such as miscalibration of pulse lengths, and intensity fluctuations, leading to over/under-rotations. In order to characterize these effects, the single qubit operations of the system were analyzed using gate set tomography (GST). GST is a method to reliably characterize a gate set of interest, inclusive of state preparation and measurement (SPAM) errors. It is implemented by applying a specific sequence of gates to a qubit, and then testing various models to see how well they fit the corresponding data set \cite{Blume:2013}. The set of gates to be analyzed was generated using the open-source pyGSTi python package developed by Sandia \cite{Sandia}. In order to detect errors from rotations around the wrong axis, it is necessary to use gate sets which include a combination of single qubit rotations \cite{Blume:2017} -- here, combinations of $\sigma_{x}\sigma_{y}$ and $\sigma_{x}\sigma_{z}$ are used. Each gate set consists of sequences comprised of combinations of $\pi/2$ single-qubit rotations, with the sequences varying in length; these gate sets are applied to a single ion initialised in the $\ket{\text{S}_{1/2},\text{m} =+1/2}$ state, followed by state-readout. The measurements were consequently analyzed using a hybrid scheme of linear inversion GST and maximum likelihood estimation available through the pyGSTi package \cite{Blume:2017, Sandia}. The GST analysis found an average gate infidelity for the $\sigma_{x}$ rotation of 0.06\%, for the $\sigma_{y}$ rotation of 0.06\%, and for the $\sigma_{z}$ rotation of 0.8\%. The relatively large error of the $\sigma_{z}$ rotation in comparison to the $\sigma_{x,y}$ rotations is predominantly due to the differing sizes of the beams used to implement the rotations. The $\sigma_{x,y}$ rotations are implemented using a broad, global beam which can illuminate the entire ion string at the same time. In contrast, the $\sigma_{z}$ rotation is achieved using a tightly-focused, single-ion addressing beam, which consequently suffers from additional instabilities due to its small beam waist (on the order of $2~\mu\mathrm{m}$).

\textit{Error during entangling operations}--  Depolarisation noise during entangling operation, in our case, can be caused by spin flips through the sideband/carrier excitation and spontaneous decay channels. Spontaneous decay of the metastable qubit state $|1\rangle$ occurs with a rate constant of $\gamma =$ 1.168(7)/s \cite{Barton2000} per qubit and additional the incoherent spin-flip processes due to unwanted excitation in the laser ion interaction have a similarly small rate constant ($ < 0.25\%$ in 5 ms). Besides this we measure entangling laser field fluctuating in intensity by $<0.3\%$, contributing to $<0.15\%$ fluctuations in the Ising interaction Hamiltonian. For the current measurements, this source of error is insignificant in comparison to the other sources, hence it will be neglected in our numerical simulations.

\section{Numerical simulation to include decoherence effects}
\label{subs:Numerics}
\begin{figure}[t]
	\centering
	\includegraphics[scale=0.55]{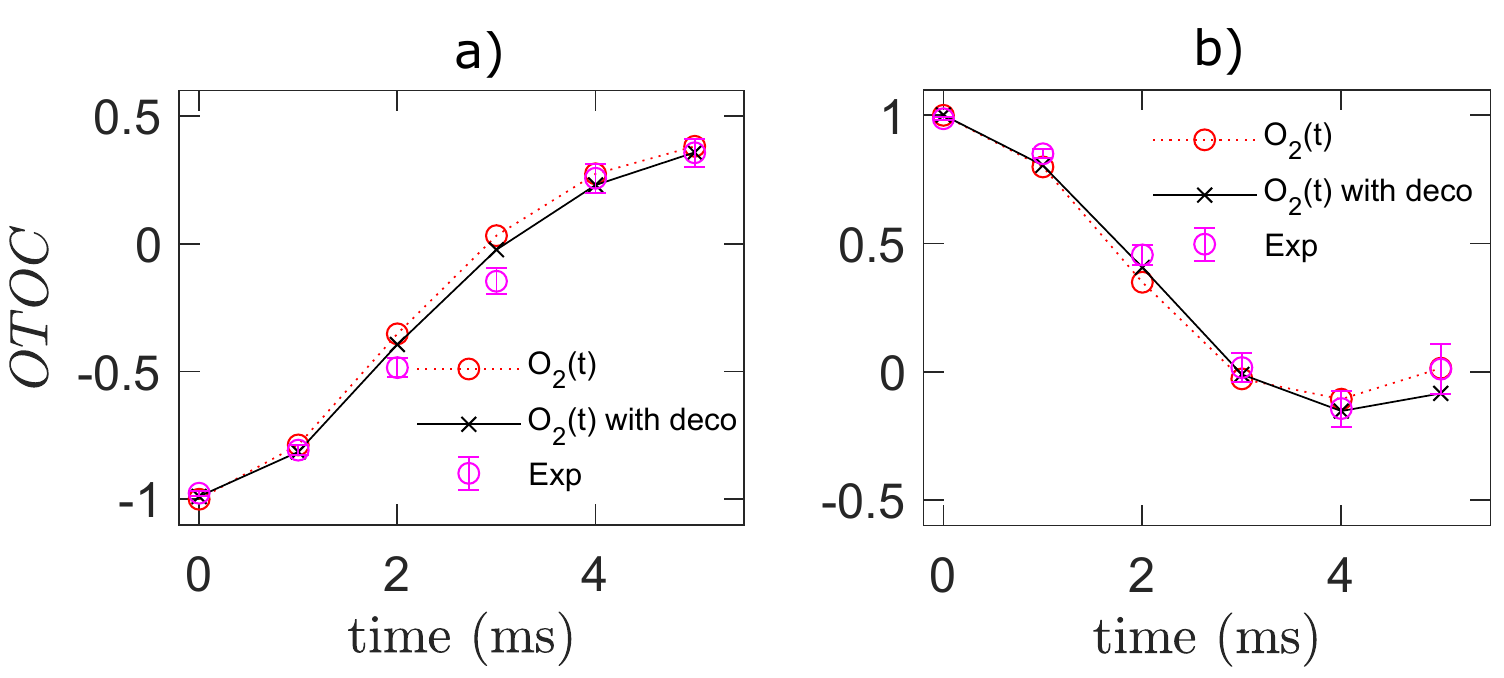}
	\caption{Numerical simulation of the protocol extracting the  OTOC $O_2(t)$ with and without decoherence effects as a function time for $\alpha=1.21$; a) $j=1$ and b) $j=2$. For comparison the experimental values of $O_2(t)$ are also shown.}
	\label{fig:SimulatedOTOC_WithDecoh_1p2}
\end{figure}
The numerical simulations were realized by propagating the Schr\"odinger equation for the different initial states $k_s$ under the Hamiltonian $H$. To include the effect of decoherence induced by shot-to-shot magnetic fluctuations, we considered that, before each projective measurement $s=1,\dots,N_m$, a global random magnetic field $B_s$  was applied for time $t$. 
The field $B_s$ has  two contributions: white noise (modeled as a discrete time sequence of normally distributed random magnetic fields with standard deviation $2\pi\times120$\,Hz and time step dt=0.0001$\,s$) 
and periodic noise 
(with amplitude $2\pi\times 90$~Hz and frequency $2\pi\times204$~Hz), whose parameters were measured independently. 
Note that our numerical simulations assume that the effect of unitary evolution and decoherence can be treated sequentially (i.e., the evolution operators), which is valid at large $B\gg J_0$.

At the theory level, we can compare the simulations described above, with simulations without decoherence of the protocol, and with a direct calculation of the OTOC \mbox{$O(t)=D^{-1}\mathrm{Tr}(W(t)V W(t)V)$}. We present an example showing the three different curves in Fig.~\ref{fig:SimulatedOTOC_WithDecoh_1p2}, showing that the main source of decoherence in our system has practically no effect on the measurement protocol.

Numerical estimations of R\'enyi entanglement entropies are based on the same approach: for each run of the experiment, we consider time evolution with a noisy magnetic field, and the projective measurement takes place after the application of a random unitary. We access then the R\'enyi entropy from such randomized measurements data~\cite{Brydges2019}. As in the experiment, we used 500 random unitaries, with 150 projective measurement per random unitary.

\section{Calibration of the spin-spin interaction}\label{subs:caliJij}

\begin{figure}[htb!]
	\centering
	\includegraphics[scale=0.4]{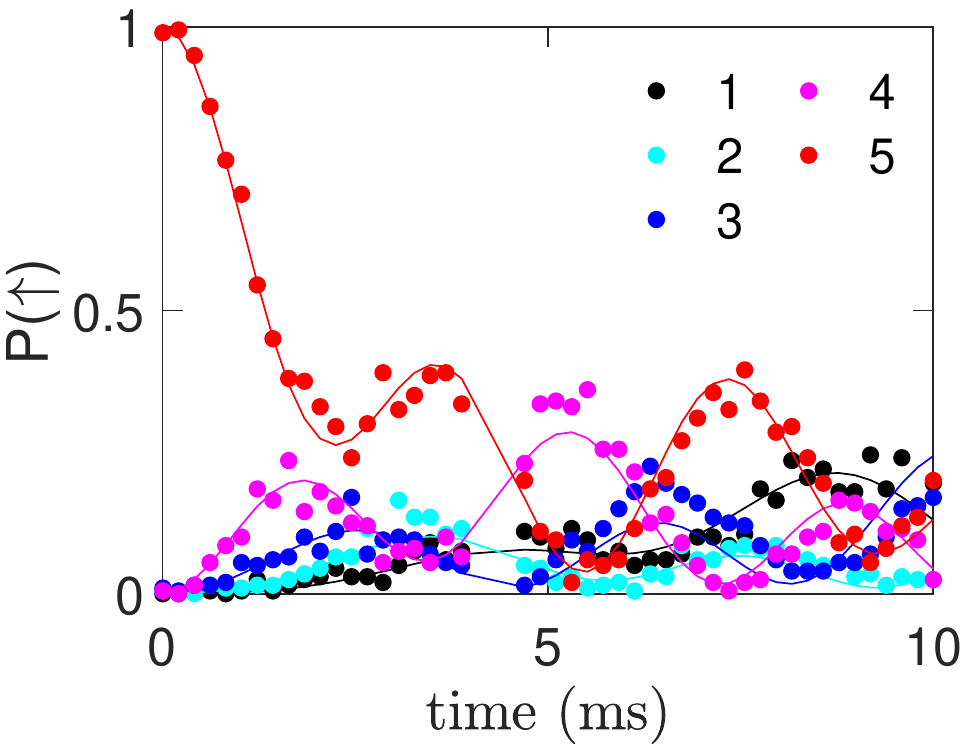}
	\includegraphics[scale=0.4]{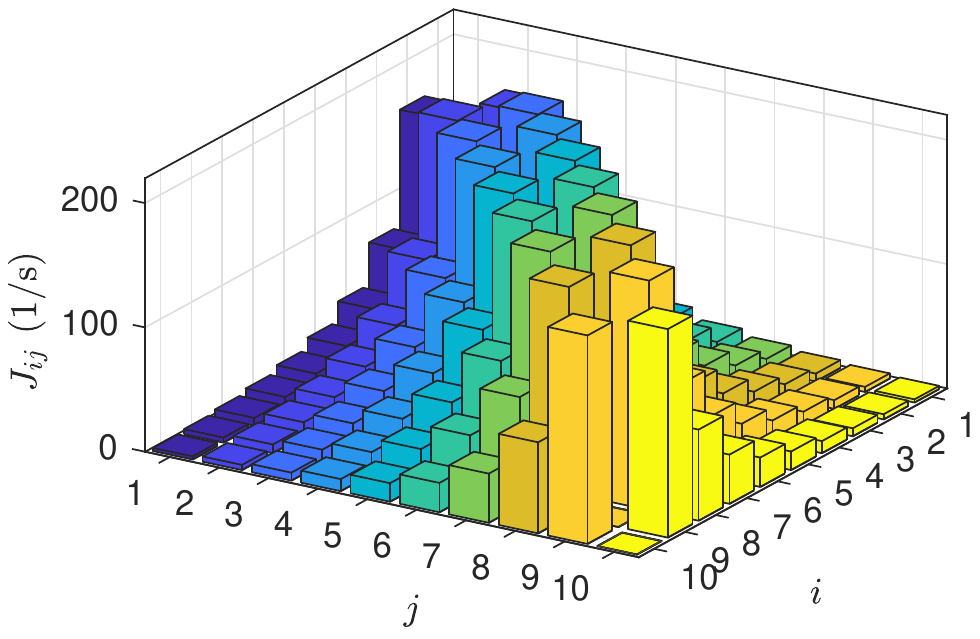}
	\caption{Spin dynamics for $\alpha=1.2$. Left: Spatial spread of excitation as a function of quench time. Initially,  spin $5$ is in the spin-up state and the remaining spins are in spin-down states.  Right: Extracted interaction matrix $J_{ij}$. In both panels, different colors represent different spins $j$.  }
	\label{fig:CorrelationSpreadalpha1p2}
\end{figure}

\begin{figure}[htb!]
	\centering
	\includegraphics[scale=0.4]{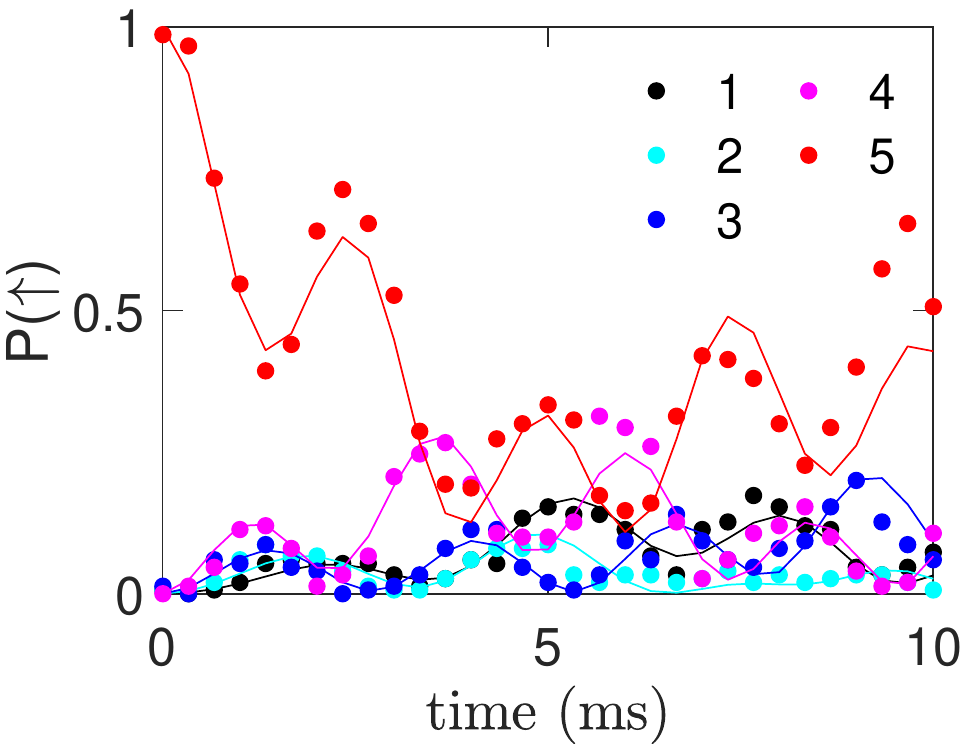}
	\includegraphics[scale=0.4]{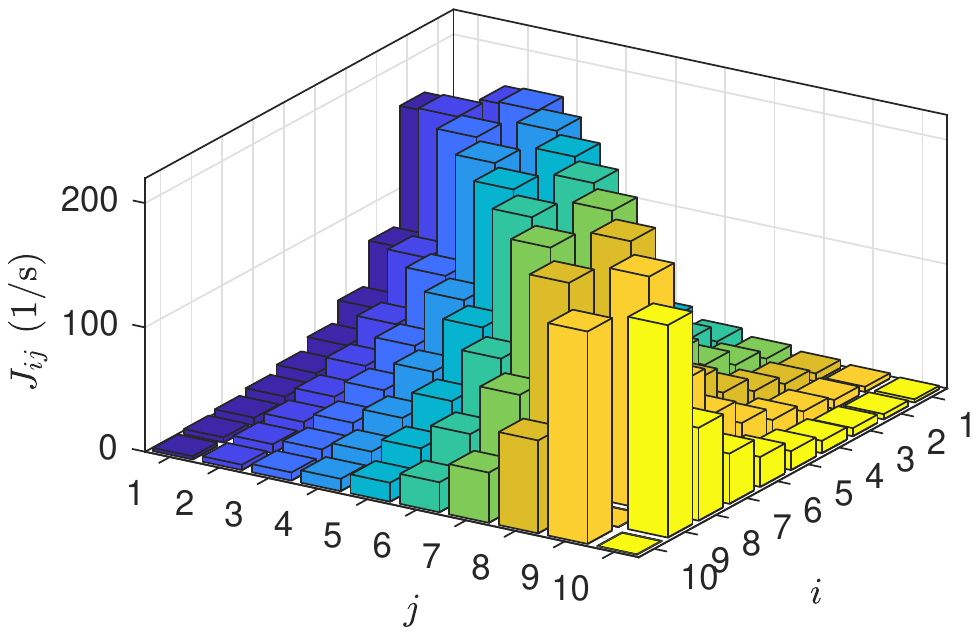}
	\caption{Spin dynamics for $\alpha=0.85$. Left: Spatial spread of excitation as a function of quench time. Initially,  spin $5$ is in the spin-up state and the remaining spins are in spin-down states.  Right: Extracted interaction matrix $J_{ij}$. In both panels, different colors represent different spins $j$.  }
	\label{fig:CorrelationSpreadalpha0p85}
\end{figure}

Here we show the excitation dynamics and its use in calibrating the spin-spin interaction. We experimentally measure the excitation spread in our 10-qubit system in the presence of the long-range interaction for two cases as discussed the paper. We start with one of the spins ($j=5$) in the spin-up state (see Ref.~\cite{Jurcevic2014} for more details). The magnetization dynamics then recorded for a variable quench time and least-squares fitting with the experimental data is performed. 
The strength of the interaction Hamiltonian is then estimated for our system with a power-law term as shown in Eq.~(2) MT  and the parameters are calculated to be $J_0 = 2\pi\times30.13$ Hz, $\alpha= 1.21$ for Fig. \ref{fig:CorrelationSpreadalpha1p2}, and $\alpha = 0.85$, $J_{0}= 2\pi \times 40.78$ Hz for Fig.~\ref{fig:CorrelationSpreadalpha0p85}. In this case, we now see that a slight mismatch with the fitted values, indicating some inconsistency between the Hamiltonian that is fitted (model Hamiltonian) and experimentally realized. The model Hamiltonian extracted here is then used in our numerical simulations presented in Fig.\ 2, MT. A small mismatch in the simulated and experimental OTOCs is thus attributed to some inconsistency between the experimentally realized and model Hamiltonian.

\section{Collapse dynamics of OTOCs in long range systems}\label{subs:CollapseFitting}

Here, we discuss scaling of the time axis $t \to t-(j-1)/v_2$ of the OTOC data, where $v_2$ denotes the velocity at which $O_2(t)$ spreads over the entire system. To extract the velocity $v_2$, we interpolate OTOCs for the intermediate time steps for spins 2--5 and extract time $t_c$, such that $O_2(t_c) = 0.5$. The data points are then fitted with a linear function $t_c=(j-1)/v_2$ to obtain $v_2$. Due to the interpolation step, giving $t_c$ from discrete time steps, the value of $v_2$ slightly depends on the interpolation method (polynomial, hyperbolic tangent fitting), resulting in an error $\Delta v_2$ of order $200\,s^{-1}$. The resulting plots are presented in Fig. (3) MT. 
 
\section{Convergence of modified OTOCs}\label{subs:Convergence} 
\begin{figure}[htb!]
	\centering
	\includegraphics[scale=0.20]{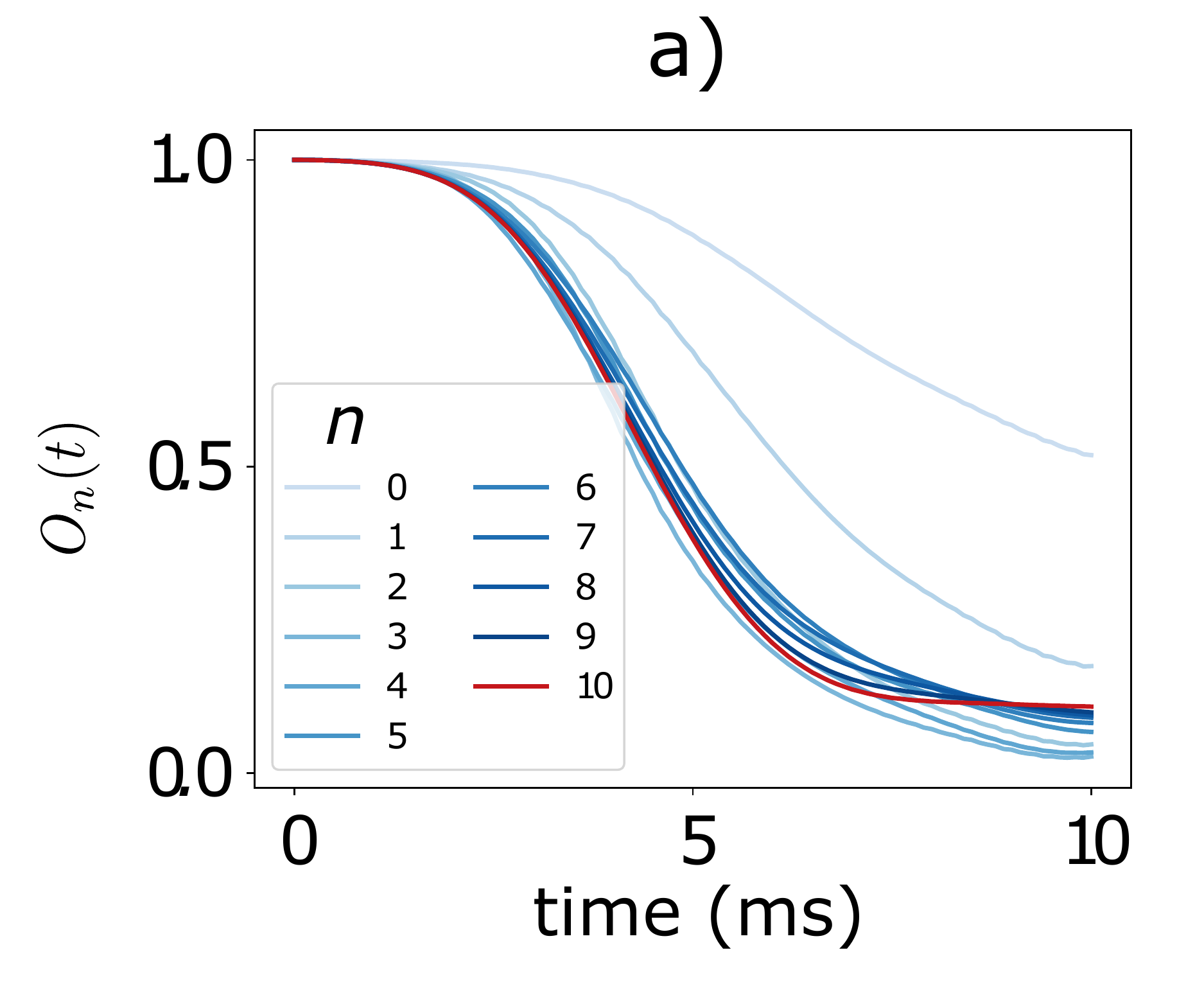}
	\includegraphics[scale=0.20]{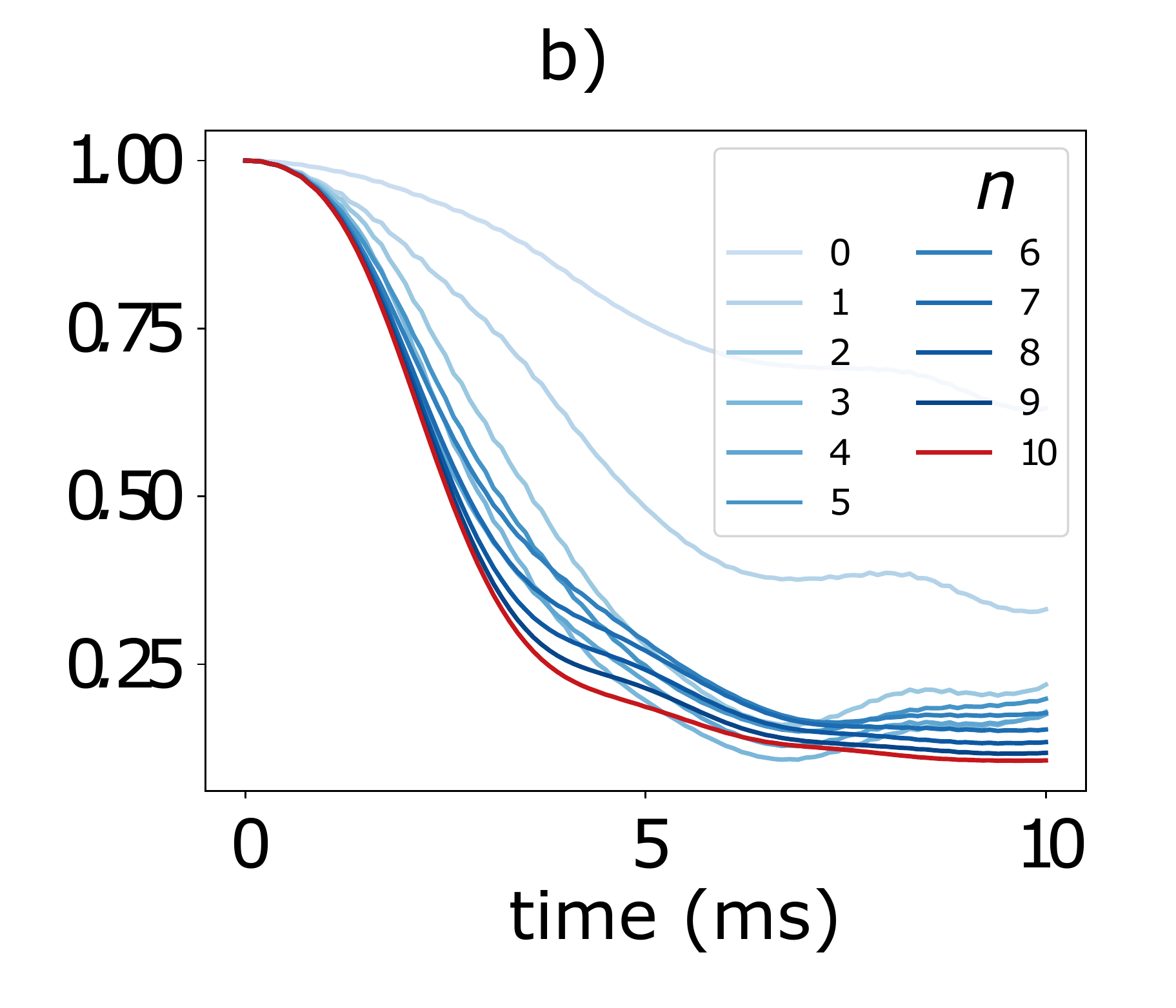}
	\caption{Comparison of modified OTOCs $O_n$ (in different shades of blue) and the exact OTOC (in red) for our spin model for a) $\alpha=1.21$ and b) $\alpha=0.85$. }
	\label{fig:ConvergenceTheory}
\end{figure}

In the main text, we discuss various orders of modified OTOCs for a non-exponential number of initial states. For example $O_0$, is defined for an initial state $\mathbf{k_0}=(0,0,....0)$. Similarly, the modified OTOCs are expanded to various orders $n$ while considering $2^n$ initial states formed from all the possible combinations of spin 0 and 1's. In this subsection, we show their convergence as a function of the order $n$ to the exact value with numerical simulations. Fig. \ref{fig:ConvergenceTheory} represents $O_n(t)$ estimated for $V=\sigma^z_1$ and $W=\sigma^x_5$ for $\alpha=1.21$ (in the left panel) and $\alpha=0.85$ (in the right panel). Parameters used in the simulations are the same as described in the main text. Notably, the convergence of modified OTOCs is asymptotic and they eventually converge to the exact OTOCs for $n=2$. This implies the measurements with 4 initial states are sufficient for the quantitative assessment of scrambling in our case. 

\end{document}